\documentclass{article}
\usepackage{bigsky2007}
\usepackage{graphicx}
\frompage{000} \topage{000}                                              
\def\rightmark{Particle Ratios on the Near and Away-Side of Jets}
\def\leftmark{J.X.Zuo et al.}

\title{Particle Ratios on the Near and Away-Side of Jets at RHIC}
\authors{
{Jiaxu Zuo$^{1,2}$ for the STAR Collaboration %
}\\[2.812mm]
{\normalsize \hspace*{-8pt}$^1$ Nuclear Physics Division, Shanghai
Institute of Applied Physics, Chinese Academy of Sciences, Shanghai 201800, P. R. China\\[0.2ex]
\hspace*{-8pt}$^2$ Physics Department, Brookhaven National Lab, 11973 Upton, NY, USA\\
}}

\abstract{ We measure the relative abundances of strange mesons,
  baryons, and anti-baryons correlated with high-$p_T$ trigger
  particles in $^{197}$Au + $^{197}$Au collisions at $\sqrt{s_{NN}}$ =
  200 GeV. Particle yields and ratios are extracted on the near-side
  and away-side of the trigger particle. The associate-particle ratios
  are studied as a function of the angle relative to the trigger
  particle azimuth $\Delta\phi$. Such studies should help elucidate
  the origin of the previously observed correlations and their strong
  modifications in Au+Au collisions relative to p+p collisions.  We
  discuss how these measurements might be related to several scenarios
  for interactions of fast partons with the medium in Au+Au
  collision.}


\begin{document}

\maketitle
\setcounter{page}{1}

\section{Introduction}\label{intro}

The observation of large collective
flows~\cite{STAR_PRL90_flow,STAR_PRL90_flow2} and
jet-quenching~\cite{STAR_PRL90_Hardtke,STAR_PRL91_highpt}
indicates that a dense medium is created in Au+Au collisions at
RHIC~\cite{STAR_NPA_whitepapers}.  Studies of two particle
azimuthal correlations have revealed detailed information about
jet interactions with this medium
\cite{STAR_PRL90_Hardtke,STAR_PRC73_2006,STAR_PRC75_2007}. These
measurements can be used to infer properties of the medium such as
it's temperature, density, and viscosity.

First measurements of di-hadron correlations show that when using
a high $p_T$ hadron to trigger on jets, in central Au+Au
collisions the away-side jet (as detected from hadrons with
$p_T>2.5$~GeV/c)
disappears~\cite{STAR_PRL90_Hardtke,STAR_PRL91_highpt}.  Later it
was shown that the remnants of the away-side jet are recovered at
lower $p_T$ values~\cite{STAR_PRL95_Fuqiang}. The distribution of
these remnants in $\Delta\phi$ is highly modified in comparison to
$p+p$ collisions: the away-side correlation is no longer peaked at
$\Delta\phi=\pi$ but instead has two peaks shifted to either side
of $\pi$~\cite{volcano}. Several scenarios have been proposed to
account for this splitting. These include: 1) the development of a
shock-wave around a fast parton traversing the
medium~\cite{Mach_Casalderrey-Solana,Mach_Stocker}. 2) the
deflection of the away-side parton as it traverses a flowing
medium~\cite{Deflect_Armesto}, and 3) the radiation of gluons at
large angles from a fast parton traversing the
medium~\cite{cherenkov_Koch}. Determining the origin of the
splitting phenomenon is experimentally challenging. Analysis of
tri-hadron correlations is being pursued as one method to distinguish
between scenarios 2 and 1 or
3~\cite{tri-hadron_Ulery,tri-hadron_Ajitanand_PHENIX}.

More information may be obtained about the interaction of fast
partons with the medium by studying the particle-type composition
of the di-hadron correlations. An increase in the ratio of baryons
to mesons has been observed in Au+Au collisions~\cite{P.Sorensen}.
The larger baryon-to-meson ratio in the in-plane direction
compared to the out-of-plane direction seems to indicate that this
increase is related to the density of the system. Models
incorporating hadron formation through coalescence of co-moving
dressed quarks successfully describe much of the observed
phenomena. By extension, one might expect a larger baryon-to-meson
ratio for intermediate $p_T$ hadrons on the away-side due to the
coalescence of quenched fragments with each other or with
constituents from the medium. Studies of the $p/\pi$ ratio show
evidence for such an effect~\cite{PHENIX_PRL91_ppbar}.

Information about the relative contribution of quarks and gluons
may also be inferred from the antibaryon-to-baryon ratio: the
fragmentation of gluon jets yields a larger antibaryon-to-baryon
ratio than the fragmentation of quark jets~\cite{quarkvsgluon}.
For this reason, if the splitting of the away-side jet is linked
to large-angle gluon radiation, then the antibaryon-to-baryon
ratio should increase at angles away from $\Delta\phi=\pi$. The
presence of these gluons may also contribute to an increase in the
baryon-to-meson ratio: a recent study found that the baryon
density is largest in collision processes involving gluons (i.e.
qg, gg, q$\overline{\mathrm{q}}$g, or
ggg)~\cite{Zhangbu_BaryonDensity}. For these reasons, measurements
of the baryon-to-meson ratio and the antibaryon-to-baryon ratio on
the near- and away-side of jets should be useful for understanding
the interaction of fast partons with the medium.

In this talk we present measurement of di-hadron correlations using
unidentified trigger hadrons and identified $K_S^0$, $\Lambda$, or
$\overline{\Lambda}$ associated partners. We study mid-central Au+Au
collisions (10\%--40\%). For this analysis, a trigger hadron is any
charged track with $3<p_T<6$~GeV/c while associated partners are taken
from $1<p_T<4$~GeV/c. We extract the baryon-to-meson and
antibaryon-to-baryon ratios of associated partners on the near-side
and away-side. We study the dependence of these ratios on $\Delta\phi$
and find that the double ratio --- the away-side particle ratios over
the near-side particle ratios --- seems to cancel dominant sources of
systematic uncertainty. We discuss our measurements and their
relationship to coalescence models, mach-cones or conical flow, and
Cherenkov radiation.

\section{Analysis method}\label{method}

\begin{figure}[htb]
\centering\mbox{ \vspace*{-8pt}
\includegraphics[width=0.7\textwidth]{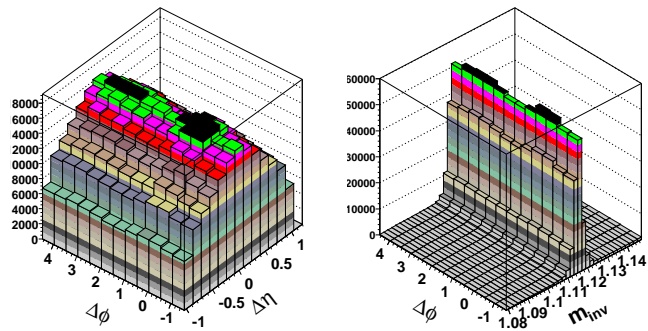}}
\vspace{-15pt} \caption[]{The $\Lambda$ plot is used as an
example. The Left Panel, 2D histogram with $\Delta\phi$ and
$\Delta\eta$. The Right Panel, 2D histogram with $\Delta\phi$ and
invariant mass. }
\label{fig1}
\end{figure}

Our analysis is carried out using charged tracks detected in the STAR
TPC~\cite{STAR_TPC}. We select trackwith pseudo-rapidity
$|\eta|<1$. $K_S^0$, $\Lambda$, and $\overline{\Lambda}$ candidates
are selected according to the invariant mass ($m_{inv}$) for track
pairs. For each trigger and associate particle pair, we fill a
3-dimensional histogram with $\Delta\phi$, $\Delta\eta$, and $m_{inv}$
($\Delta\phi = \phi_{trig} - \phi_{asso}$ and $\Delta\eta =
\eta_{trig} - \eta_{asso}$). We then sum over entries in a given
$\Delta\eta$ range (Fig.~\ref{fig1} left panel). Then we fit the
$m_{inv}$ distribution in each $\Delta\phi$ bin (Fig.~\ref{fig1} right
panel) to get the yield $dN/d\Delta\phi$ for $K_{S}^{0}$, $\Lambda$
and $\bar{\Lambda}$ as a function of $\Delta\phi$. The same procedure
is carried out on a mixed event sample to obtain a background
distribution used to correct for imperfect detector acceptance. These
effects are small because of the 2$\pi$ azimuthal coverage of the STAR
TPC. We scale the real event $dN/d\Delta\phi$ distribution by the
normalized mixed event distribution. Finally, a $v_{2}$ modulated
background distribution is subtracted from the corrected
$dN/d\Delta\phi$ distribution.  This subtraction accounts for the
correlations from $v_2$ and the combinatorial background. The level of
the background is adjusted so that the subtracted $dN/d\Delta\phi$
distribution has zero-yield at the minimum
(ZYAM)~\cite{PHENIX_PRL97_ZYAM} or zero-yield at $\Delta\phi=1$
(ZYA1). Efficiency corrections for the associated particles are then
applied to the resulting $dN/d\Delta\phi$ distribution. The following
form is used to describe the $v_2$ modulated combinatorial
background~\cite{STAR_PRL90_flow,PHENIX_PRL97_ZYAM}:
$B(\Delta\phi)=b0(1+2\langle v_{2}^{A}\times
v_{2}^{B}\rangle\cos(2\Delta\phi))$.

The background subtraction is the source of three major systematic
uncertainties: 1) uncertainty in the value of $v_2$ and $v_2$
fluctuations~\cite{STAR_v2fluctuation} ($\langle v_2^A\times
v_2^B\rangle \ne \langle v_{2}^{A}\rangle\times\langle
v_{2}^{B}\rangle$), 2) uncertainty in the assumption that the
correlations can be factorized into a jet-like component and a
combinatorial background (the two-component model), and 3) the
model dependent assumption that the $dN/d\Delta\phi$ distribution
should have zero-yield at some specified $\Delta\phi$ value (ZYAM,
ZYA1 etc.). These uncertainties lead to large systematic errors in
our analysis. The nominal $v_2$ for the charged hadron trigger
particle is taken as the average of $v_2$ from an event plane
analysis ($v_{2}\{EP\}$) and $v_2$ from a 4-particle cumulant
analysis ($v_{2}\{4\}$)~\cite{STAR_PRC66_v24}. $v_{2}\{EP\}$ and
$v_{2}\{4\}$ are used as upper and lower bounds respectively for
the allowed $v_2$. For the associated $K_{S}^{0}$, $\Lambda$ and
$\bar{\Lambda}$~\cite{STAR_PRL95_Flow_KL}, the same procedure is
followed except that $v_{2}\{LYZ\}$~\cite{LYZ} is used as the
lower bound instead of
$v_{2}\{4\}$~\cite{STAR_PRL89_KsLv2,STAR_PRL92_partdepend}.

\section{Results}\label{result}

\begin{figure}[htb]
\vspace{-5pt}
\resizebox{0.5\textwidth}{!}{\includegraphics{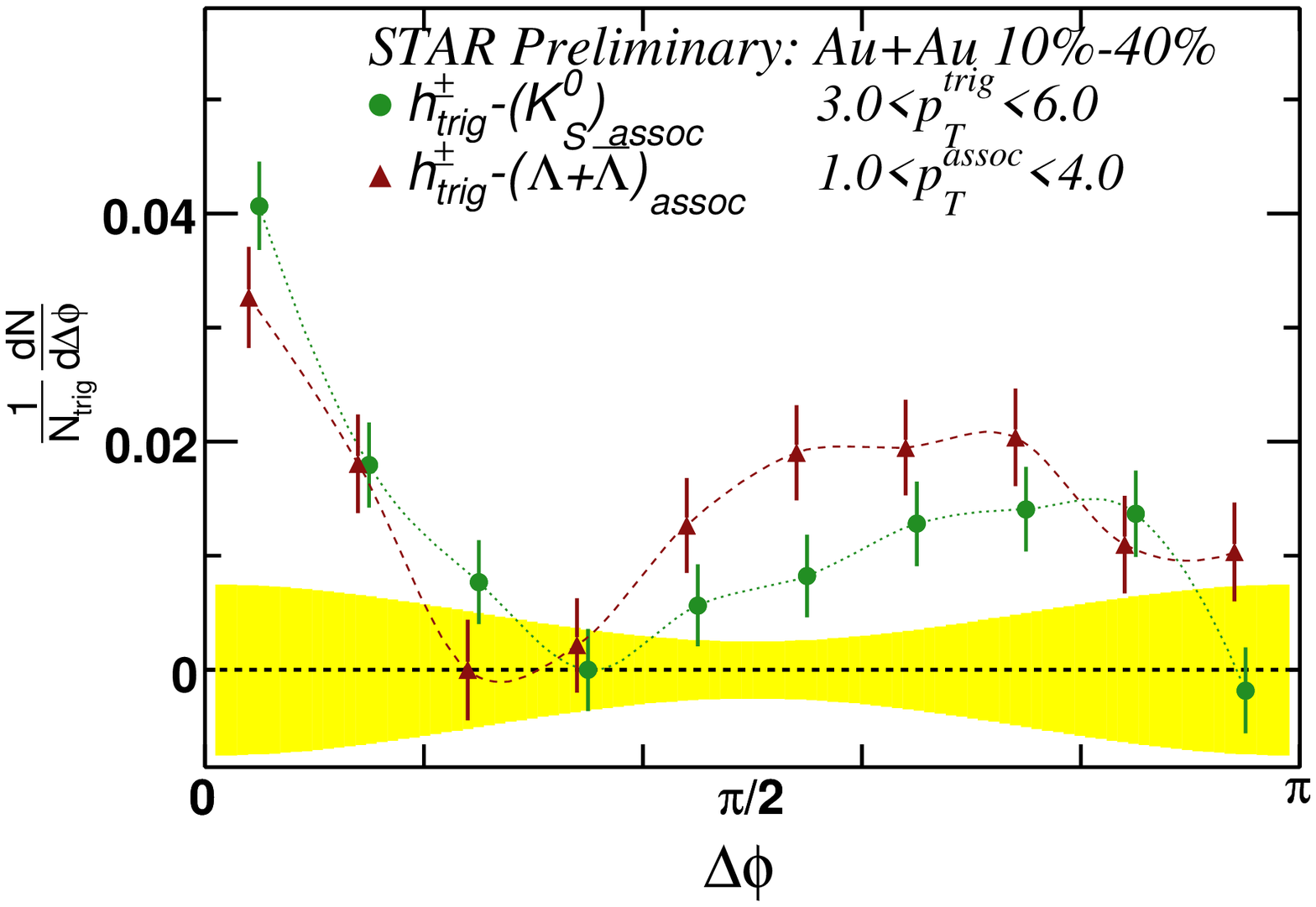}}
\resizebox{0.5\textwidth}{!}{\includegraphics{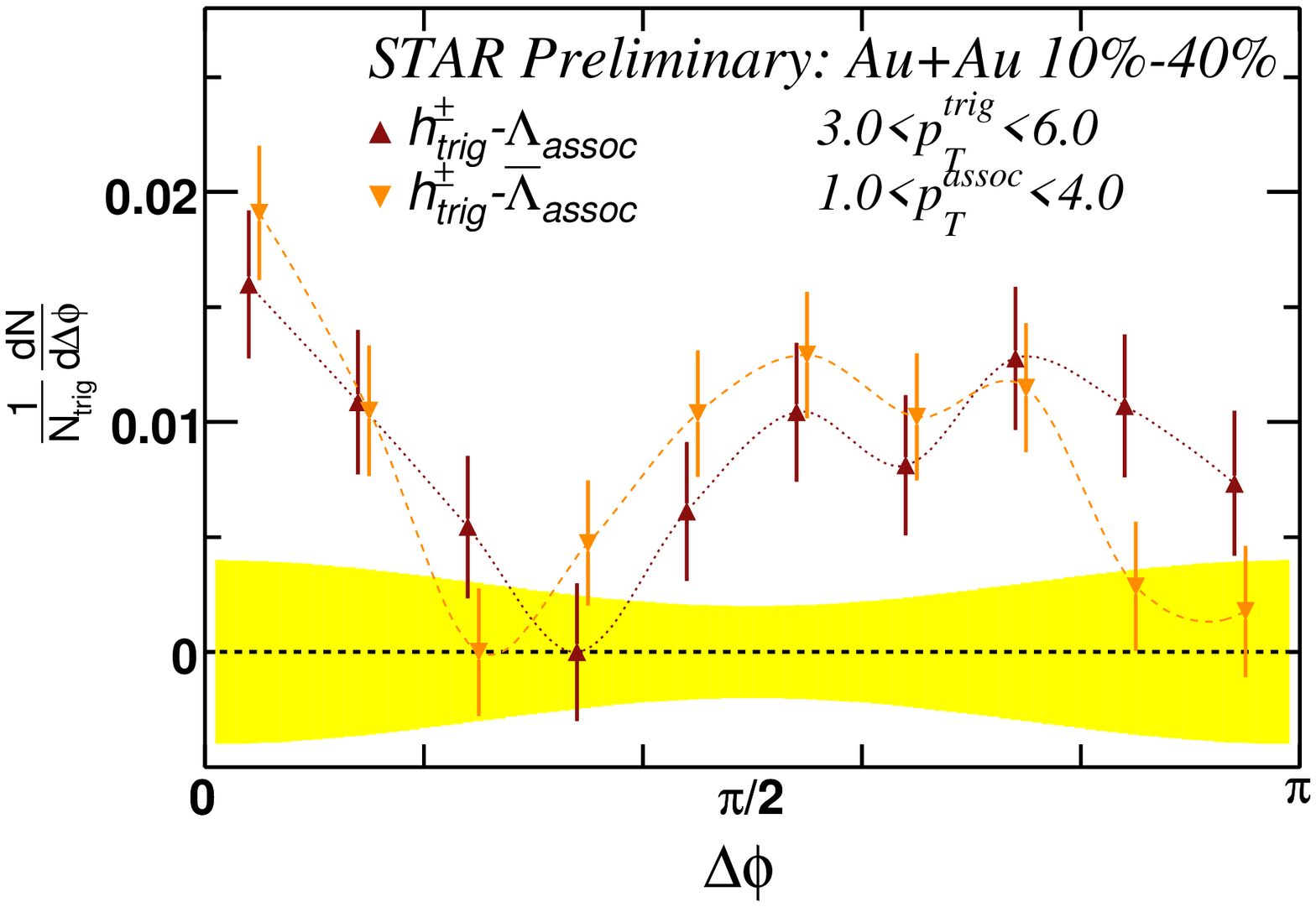}}
\caption[]{Hadron-$K_{S}^{0}$, -$\Lambda$+$\overline{\Lambda}$ (left panel)
  and hadron-$\Lambda$, -$\overline{\Lambda}$ (right panel) correlation
  function in the centrality bin $10\%-40\%$ in $^{197}$Au +
  $^{197}$Au collisions at $\sqrt{s_{NN}}$ = 200 GeV. The trigger
  particles $p_{T}$ range is $3.0<p_{T}<6.0$; the associate
  $K_{S}^{0}$, $\Lambda$, or $\overline{\Lambda}$ particles $p_{T}$ range
  is $1.0<p_{T}<4.0$. The yellow band around the zero is the
  systematic errors. }
\label{fig2}
\end{figure}

The acceptance, efficiency, and background subtracted di-hadron
$dN/d\Delta\phi$ distributions are shown in Fig.~\ref{fig2}. The
data are from the 10\%--40\% centrality interval of
$\sqrt{s_{NN}}$ = 200 GeV Au+Au collisions. All data is from the
$\eta$ window $|\eta|<1.0$. The left panel shows the
hadron-$K_S^0$, and the hadron-($\Lambda+\overline{\Lambda}$)
$dN/d\Delta\phi$ distributions. The right panel shows the
hadron-$\Lambda$ and hadron-$\overline{\Lambda}$ correlations
separately. The yellow band around zero represents the systematic
uncertainties. For all particle combinations a strong correlation
is seen on the near-side of the charged hadron trigger
($\Delta\phi<1.1$) as would be expected from fragmentation of a
fast parton or jet. The correlation structure on the away-side of
the trigger hadron is very broad and may even exhibit a minimum at
$\Delta\phi=\pi$ where typically a maximum would exist. These
features are similar to those already observed for unidentified
di-hadron distributions which have much better
statistics~\cite{PHENIX_PRL97_ZYAM}.

\begin{table}[hb]
\vspace*{-12pt} \caption[]{Ratios of identified partner particles
  ($1<p_{T}<4$~GeV/c) associated with a charged hadron trigger
  particle ($p_{T}>3.0$~GeV/c).}
\label{tab1}\vspace*{-10pt}
\begin{center}
\begin{tabular}{c|c|c}
\hline\\[-10pt]
Particle Ratios & Near-Side (Stat. Sys.) & Away-Side (Stat. Sys.)\\
\hline\\[-10pt]
$(\Lambda+\overline{\Lambda})/K_{S}^{0}$ & $0.765 \pm 0.120 \pm 0.175$
& $1.71 \pm 0.321 \pm 0.589$\\
$\overline{\Lambda}/\Lambda$ & $0.916 \pm 0.200 \pm 0.200$ &
$0.894 \pm 0.173 \pm 0.368$\\
\hline
\end{tabular}
\end{center}
\end{table}

We extract the conditional yields of identified $K_{S}^{0}$, $\Lambda$
and $\overline{\Lambda}$ particles on the near-side
($0.<\Delta\phi<0.35\pi$) and away-side ($0.35\pi<\Delta\phi<\pi$) of
the trigger hadron. In Table~\ref{tab1} we present the resulting
$(\Lambda+\overline{\Lambda})/K_S^0$ and the
$\overline{\Lambda}/\Lambda$ ratios on the near- and away-side along
with the systematic and statistical errors.

\begin{figure}[htb]
\centering\mbox{ \vspace*{-8pt}
\includegraphics[width=0.80\textwidth]{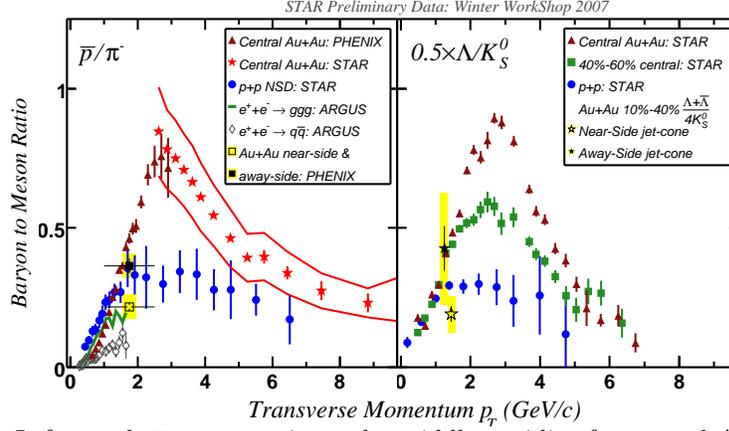}}
\vspace*{-15pt} \caption[]{ Left panel: $\overline{p}$ to $\pi^{-}$
ratio at the middle rapidity for central Au+Au and p+p collisions
at 200 GeV. The Measurement of the proton to pion ratio made for
particles associated with a trigger hadron ($p_{T}>2.5$) are also
shown. Right panel: $\Lambda$ to $K_{S}^{0}$ ratio in central
Au+Au, mid-peripheral Au+Au and minimum-bias p+p collisions. The
Measurement of the $\Lambda$ to $K_{S}^{0}$ ratio made for
particles associated with a trigger hadron ($p_{T}>3.0$) are shown.
Values are scaled by 0.5. In the plots, the yellow band is the
systematic errors. } \label{fig4}
\end{figure}

In Fig.~\ref{fig4} we compare our results for the
$(\Lambda+\overline{\Lambda})/K_S^0$ ratio to other measurements of
the baryon-to-meson ratio. The left panel shows the
$\overline{p}/\pi^-$ ratio measured in $e^++e^-$~\cite{argus},
$p+p$~\cite{STAR_PLB637_Identified_pp}, and
$Au+Au$~\cite{PHENIX_PRL91_ppbar,STAR_PRL97_Identified_AuAu}
collisions (these measurements are not conditional yields). The right
panel shows the $\Lambda/K^0_S$ ratio for $p+p$, mid-peripheral
$Au+Au$~\cite{STAR_str}, and central $Au+Au$ collisions scaled by
0.5. In central Au+Au collisions, $\overline{p}/\pi^-$ reaches a
maximum value of nearly 1 at $p_T\approx3$~GeV/c. The measurements of
the $\overline{p}/\pi^-$ ratio made for particles associated with a
trigger hadron ($p_{T}>2.5$) from PHENIX are also shown in the left
panel while our results from Tab.~\ref{tab1} are shown in the right
panel. We find that both STAR and PHENIX measurements are consistent
with a larger baryon-to-meson ratio on the away-side than on the
near-side. In addition, on the near-side the baryon-to-meson ratio is
closer to values measured in p+p collisions while on the away-side the
ratio is closer to that measured in central or mid-central Au+Au
collisions. This observation may indicate that the larger density of
matter traversed by the away-side jet leads to an enhancement in
baryon production. Such an effect is expected if the baryon
enhancement in the intermediate $p_T$ region observed in Au+Au
collisions is due to multi-parton interactions such as gluon
junction~\cite{gluon_junction} or quark
coalescence~\cite{coalescence}.

Table~\ref{tab1} gives the particle ratios integrated over wide
regions of $\Delta\phi$ (\textit{i.e.} the near- and away-side).  More
information can be obtained from the distributions in Fig.~\ref{fig2}
by examining how the ratios depend on $\Delta\phi$: \textit{e.g.} the
ratios of the conditional yields on the away-side can help us better
understand the source of the correlations that appear at large angles
away from $\Delta\phi=\pi$. It has been speculated that the enhanced
correlations at wide angles may be related to large angle gluon
radiation~\cite{cherenkov_Koch,Gluonradiation_PLB78_Vitev}, deflection
of the away-side jet by the flowing medium~\cite{Deflect_Armesto}, or
a shock wave that is induced in the medium by a fast moving
parton~\cite{Mach_Casalderrey-Solana,Mach_Stocker}. We expect the
dependence of the particle ratios on $\Delta\phi$ to differ in the
three above scenarios: \textit{e.g.} gluons radiated at large angles
may lead to a larger antibaryon-to-baryon ratio in that region. A
recent study also found that the presence of these gluons may also
lead to an enhanced baryon-to-meson
ratio~\cite{Zhangbu_BaryonDensity}. Alternatively, the higher density
that would be associated with a shock-wave could lead to an increase
in the baryon-to-meson ratio via coalescence of co-moving partons. It
has also been argued that since a shock wave wave should be moving at
the speed of sound in the medium, the particles produced from such a
shock should not be very fast particles. For a slow particle to
satisfy the $p_T$ cut in our analysis it would have to be massive. For
this reason, one might expect the correlation at large angles to have
a larger number of massive particles and consequently a larger
baryon-to-meson ratio. Detailed calculations of particle ratios from
the above scenarios have not been made but are being
pursued~\cite{ourNewPaper}.

\begin{figure}[htb]
\vspace{-5pt}
\resizebox{0.5\textwidth}{!}{\includegraphics{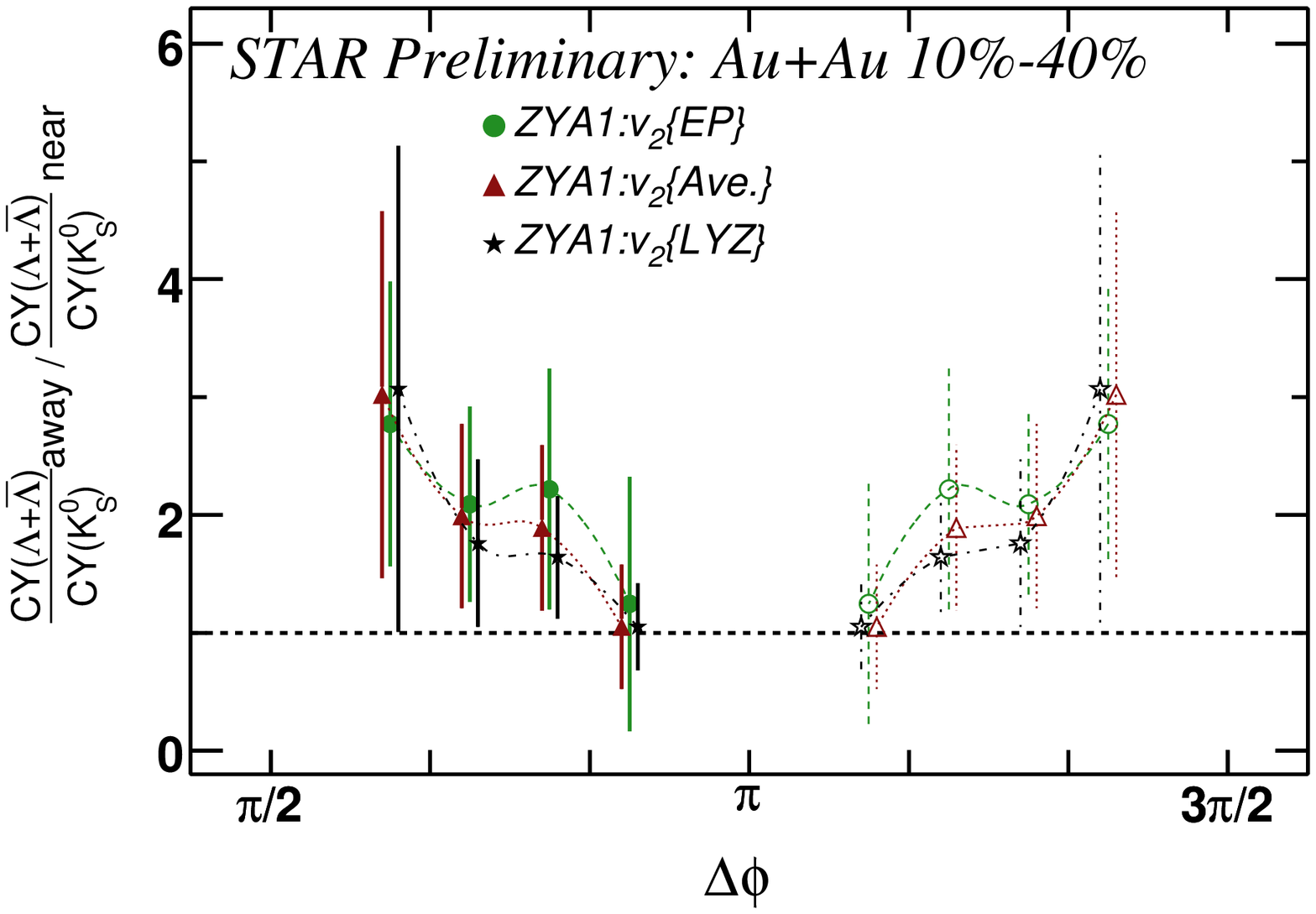}}
\resizebox{0.5\textwidth}{!}{\includegraphics{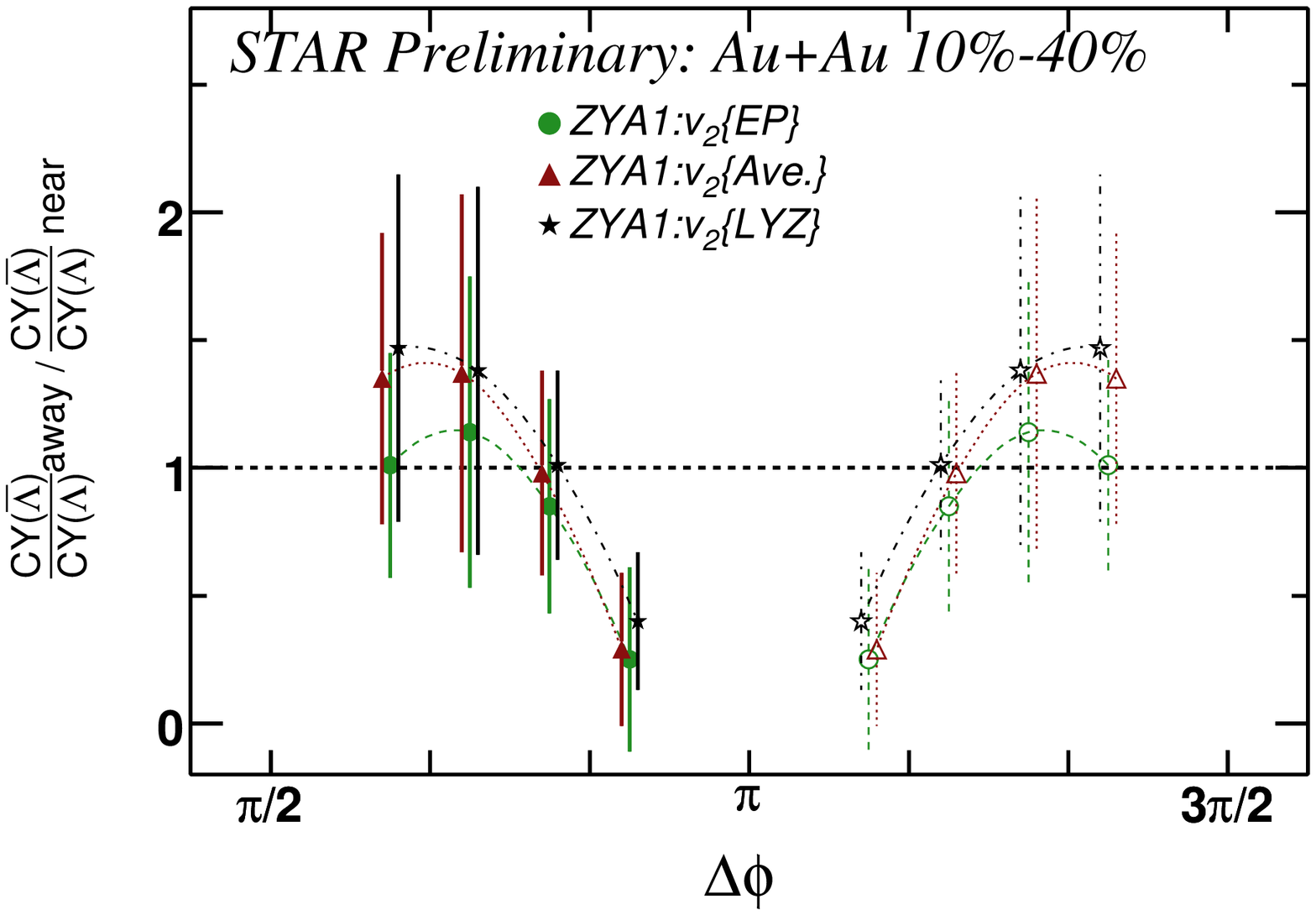}}
\caption[]{Left panel: The baryon-to-meson ratio on the away-side
  vs. $\Delta\phi$ scaled by the same ratio in the near-side
  jet-cone. Data are for $10\%-40\%$ Au+Au collisions at 200 GeV. This
  double ratio appears to be insensitive to the background subtraction
  method. Right panel: the same for $\overline{\Lambda}/\Lambda$.
} \label{fig5}
\end{figure}

Fig.~\ref{fig5} shows the particle ratios (ratios of the conditional
yields) on the away-side as a function of $\Delta\phi$. The ratios are
normalized by the corresponding ratio measured in the near-side
jet-cone so that unity corresponds to the case where the away-side
particle composition is the same as that in the near-side jet cone. We
find that this double ratio is largely independent of the elliptic
flow used in the background subtraction indicating that such an
analysis is able to reduce systematic uncertainties. The left panel
shows the baryon-to-meson, awayside-to-nearside double ratio and the
right panel shows the antibaryon-to-baryon, awayside-to-nearside
double ratio. In both cases the data from $\Delta\phi<\pi$ (closed
symbols) has been reflected to $\Delta\phi>\pi$ (open symbols). The
uncertainty on both measurements remains large and precludes strong
conclusions about the shape or magnitude of the ratios. We see some
indication that both the $\overline{\Lambda}/\Lambda$ and
$(\overline{\Lambda}+\Lambda)/K_S^0$ ratios are large at large angles
than they are at $\Delta\phi=\pi$. This would be consistent, for
example, with gluon radiation at large angles as discussed above.

The uncertainty in our measurements can be reduced in several
ways. First, a more precise determination of $\langle v_2^A\times
v_2^B\rangle$ can eliminate that source of error entirely. These
measurements are currently being pursued in STAR. Second, greater
statistics can reduce the uncertainty in the background normalization
via ZYAM or ZYA1. Greater statistics will become available in upcoming
runs at RHIC so we expect to be able to improve the precision of our
measurements in the near future. Other uncertainties in our analysis
due to the ZYAM assumption and the two-component, jet+flow background
model may be un-reducible.

\section{Summary}\label{sum}

We measured di-hadron azimuthal angle correlations in Au+Au collisions
at $\sqrt{s_{NN}}$ = 200 GeV. Charged hadrons ($3.0<p_{T}<6.0GeV/c$)
are used as the trigger particle; $K_{S}^{0}$s, $\Lambda$s and
$\overline{\Lambda}$s ($1.0<p_{T}<4.0$ GeV/c) are used as the
associated particles. A correlation is observed between the trigger
hadrons and $K_{S}^{0}$s, $\Lambda$s and $\overline{\Lambda}$s. We
extracted the conditional yields of identified associate particles on
the near- and away-side of the jet trigger and calculated the near and
away-side particle ratios.  The systematic uncertainty from $v_{2}$
and the background normalization are large. These uncertainties can be
reduced with more data to reduce the error on the level of the
background and a better understanding of $v_{2}$ to reduce uncertainty
on the shape of the background. Both STAR and PHENIX results are
consistent with a larger baryon-to-meson ratio on the away-side than
the near-side. We studied the shape of away-side particle ratios and
find that this shape is insensitive to several sources of systematic
uncertainty. These measurements should help elucidate how fast partons
interact with the matter created in Au+Au collisions at RHIC.

\textbf{Acknowledgments:} This work is completed with P. Sorensen. We
thank Z. Xu, A. H. Tang and the STAR Group at BNL for enlightening discussions.

\vfill\eject
\end{document}